\documentclass[twocolumn,preprintnumbers,nofootinbib,10pt,prl]{revtex4-1}

\usepackage{graphicx}
\usepackage{amsmath}
\usepackage{amssymb}
\usepackage{verbatim}
\usepackage{mathrsfs}
\usepackage{array}
\usepackage{layout}
\usepackage{textcomp}
\usepackage{latexsym}
\usepackage{slashed}
\usepackage{rotating}
\usepackage{hyperref}
\usepackage{booktabs}
\usepackage{multirow}
\usepackage[usenames,dvipsnames]{color}

\newcommand{\Eq}{&=&}

\newcommand{\white}[1]{{\color[rgb]{1,1,1} #1}}

\newcommand{\hs}[1]{{\hspace{#1}}}
\newcommand{\vs}[1]{{\vspace{#1}}}
\newcommand{\tf}[1]{{\textsf{#1}^{}}}
\newcommand{\tx}[1]{{\text{#1}^{}}}
\newcommand{\nn}{\nonumber\\}

\newcommand{\sx}[2]{{\scalebox{#1}{#2}}}
\newcommand{\dd}{{\mathrm{d}}}
\newcommand{\XXXNN}
{X\hs{-0.03cm}X\hs{-0.03cm}X \hs{-0.08cm} \to \hs{-0.08cm} \bar{N}\hs{-0.03cm}\bar{N}}
\newcommand{\XXNXN}
{X\hs{-0.03cm}X\hs{-0.03cm}N \hs{-0.08cm} \to \hs{-0.08cm} \bar{X}\hs{-0.03cm}\bar{N}}
\newcommand{\XNNXX}
{X\hs{-0.03cm}N\hs{-0.03cm}N \hs{-0.08cm} \to \hs{-0.08cm} \bar{X}\hs{-0.03cm}\bar{X}}
\newcommand{\NNXX}
{N\hs{-0.03cm}\bar{N} \hs{-0.08cm} \to \hs{-0.08cm} X\hs{-0.03cm}\bar{X}}
\newcommand{\XXNN}
{X\hs{-0.03cm}\bar{X} \hs{-0.08cm} \to \hs{-0.08cm} N\hs{-0.03cm}\bar{N}}
\newcommand{\NNXXX}
{\bar{N}\hs{-0.03cm}\bar{N} \hs{-0.08cm} \to \hs{-0.08cm} X\hs{-0.03cm}X\hs{-0.03cm}X}
\newcommand{\XXNNNN}
{X\hs{-0.03cm}\bar{X} \hs{-0.08cm} \to \hs{-0.08cm} N\hs{-0.03cm}\bar{N}\hs{-0.03cm}N\hs{-0.03cm}\bar{N}}
\newcommand{\xfo}{x^{}_\tf{f.o.}}
\newcommand{\xfi}{x^{}_\tf{f.i.}}
\newcommand{\xkd}{x^{}_\tf{k.d.}}
\newcommand{\gD}{g^{}_\tf{D}\hs{-0.06cm}}

\newcommand{\XXXNbNa}{X\hs{-0.03cm}X\hs{-0.03cm}X \hs{-0.08cm} \to \hs{-0.08cm} \bar{N}^{}_2\hs{-0.01cm}N^{}_1}
\newcommand{\XXNbXNa}{X\hs{-0.03cm}X\hs{-0.03cm}N^{}_2 \hs{-0.08cm} \to \hs{-0.08cm} \bar{X}\hs{-0.02cm}N^{}_1}
\newcommand{\XXNaXNb}{X\hs{-0.03cm}X\hs{-0.03cm}\bar{N}^{}_1 \hs{-0.08cm} \to \hs{-0.08cm} 
\bar{X}\hs{-0.02cm}\bar{N}^{}_2}
\newcommand{\XNbNaXX}{X\hs{-0.03cm}N^{}_2\hs{-0.01cm}\bar{N}^{}_1 \hs{-0.08cm} \to \hs{-0.08cm} \bar{X}\hs{-0.02cm}\bar{X}}
\newcommand{\XXNcNc}{X\hs{-0.03cm}\bar{X} \hs{-0.08cm} \to \hs{-0.08cm} N^{}_{1,2}\hs{-0.01cm}\bar{N}^{}_{1,2}}
\newcommand{\NcNcXX}{N^{}_{1,2}\hs{-0.01cm}\bar{N}^{}_{1,2} \hs{-0.08cm} \to \hs{-0.08cm} X\hs{-0.03cm}\bar{X}}
\newcommand{\NaNaXX}{N^{}_1\hs{-0.01cm}\bar{N}^{}_1 \hs{-0.08cm} \to \hs{-0.08cm} X\hs{-0.03cm}\bar{X}}
\newcommand{\XXNaNa}{X\hs{-0.03cm}\bar{X} \hs{-0.08cm} \to \hs{-0.08cm} N^{}_1\hs{-0.01cm}\bar{N}^{}_1}

\newcommand{\NaNaNbNb}{N^{}_1\hs{-0.01cm}\bar{N}^{}_1 \hs{-0.08cm} \to \hs{-0.08cm} N^{}_2\hs{-0.01cm}\bar{N}^{}_2}
\newcommand{\NbNbNaNa}{N^{}_2\hs{-0.01cm}\bar{N}^{}_2 \hs{-0.08cm} \to \hs{-0.08cm} N^{}_1\hs{-0.01cm}\bar{N}^{}_1}

\begin{document}

\vspace*{-30mm}
\font\mini=cmr10 at 0.8pt

\title{Reshuffled strongly interacting massive particle dark matter}

\author{Shu-Yu\,\,Ho$^{1}$}\email{phyhunter@kias.re.kr}
\author{Pyungwon\,\,Ko$^{1}$}\email{pko@kias.re.kr}
\author{Chih-Ting\,\,Lu$^{1,2}$}\email{timluyu@kias.re.kr\\}

\affiliation{\vspace{4pt} 
${}^{1}$Korea Institute for Advanced Study, Seoul 02455, Republic of Korea \vspace{3pt} \\
${}^{2}$Department of Physics and Institute of Theoretical Physics, Nanjing Normal University, Nanjing, 210023, China
}

\preprint{KIAS-P21025}

\begin{abstract}
In this work, we reanalyze the multi-component strongly interacting massive particle (mSIMP) scenario using an effective operator approach.\,\,As in the single-component SIMP case, the total relic abundance of mSIMP dark matter (DM) is determined by the coupling strengths of $3 \to 2$ processes achieved by a five-point effective operator.\,\,Intriguingly, we notice that there is an irreducible $2 \to 2$ process induced by the corresponding five-point interaction in the dark sector, which would reshuffle the mass densities of SIMP DM after the chemical freeze-out.\,\,We dub this DM scenario as reshuffled SIMP ($r$SIMP).\,\,Given this observation, we then numerically solve the coupled Boltzmann equations including the $3 \to 2$ and $2 \to 2$ processes to get the correct yields of $r$SIMP DM.\,\,It turns out that the masses of $r$SIMP DM must be nearly degenerate for them to contribute sizable abundances.\,On the other hand, we also introduce effective operators to bridge the dark sector and visible sector via a vector portal coupling.\,\,Notably, we find that the reshuffled mechanism in the $r$SIMP scenario is sensitive to the size of the DM self-interacting cross section.
\end{abstract}

\maketitle


\section{Introduction}

One of the greatest mysteries in cosmology is the nature and origin of dark matter (DM), which constitutes about 27\% of the energy budget in the universe.\,${}^{}$The DM can be produced from the thermal reservoir in the early universe.\,${}^{}$Thus, its relic density may be insensitive to initial conditions.\,${}^{}$The most fashionable instance of thermal DM is weakly interacting massive particles (WIMP)\,\cite{Lee:1977ua}, where the relic abundance is determined by annihilation cross sections of DM pairs into the standard model (SM) particles.\,\,Besides, strongly interacting massive particles (SIMP)\,\cite{Hochberg:2014dra} and elastically decoupling relics (ELDER)\,\cite{Kuflik:2015isi} are thermal DM alternatives that have drawn attention due to their novel dynamics.\,\,In these two paradigms, the DM abundances are set by annihilation cross sections of DM number-changing processes and by elastic scattering rates of DM with SM particles, respectively.

\renewcommand{\arraystretch}{1.5}
\begin{table}[b!]
\vs{-0.3cm}
\begin{tabular}{|c|c|c|c|}
\hline
\,Class\, & \,Condition\, 
\\\hline
WIMP & \,\,$\Gamma^{}_\tf{el} \gg \Gamma^{}_\tf{ann}  > H^{}_\tf{DM}$
\\\hline
SIMP & \,\,$\Gamma^{}_\tf{el} > \Gamma^{}_{3 \to 2} \gg \Gamma^{}_\tf{ann} > H^{}_\tf{DM}$\,
\\\hline
\,ELDER\, & \,\,$\Gamma^{}_{3 \to 2} > \Gamma^{}_\tf{el} \gg \Gamma^{}_\tf{ann} > H^{}_\tf{DM}$\, 
\\\hline
\end{tabular}
\caption{The taxonomy of thermally-produced DM, where $\Gamma^{}_\tf{el}$ is typically larger than $\Gamma^{}_\tf{ann}$ because the number density of SM particles, $n^{}_\tf{SM}$, dominates over the number density of DM, $n^{}_\tf{DM}$, much before the matter-radiation equality, and the strength of the elastic scattering cross section $\langle \sigma^{}_\tf{el} v \rangle$ is similar to that of the annihilation cross section $\langle \sigma^{}_\tf{ann} v \rangle$.}
\label{tab:taxonomy}
\vs{-0.5cm}
\end{table}

\begin{figure}[t!]
\begin{center}
\includegraphics[width=0.45\textwidth]{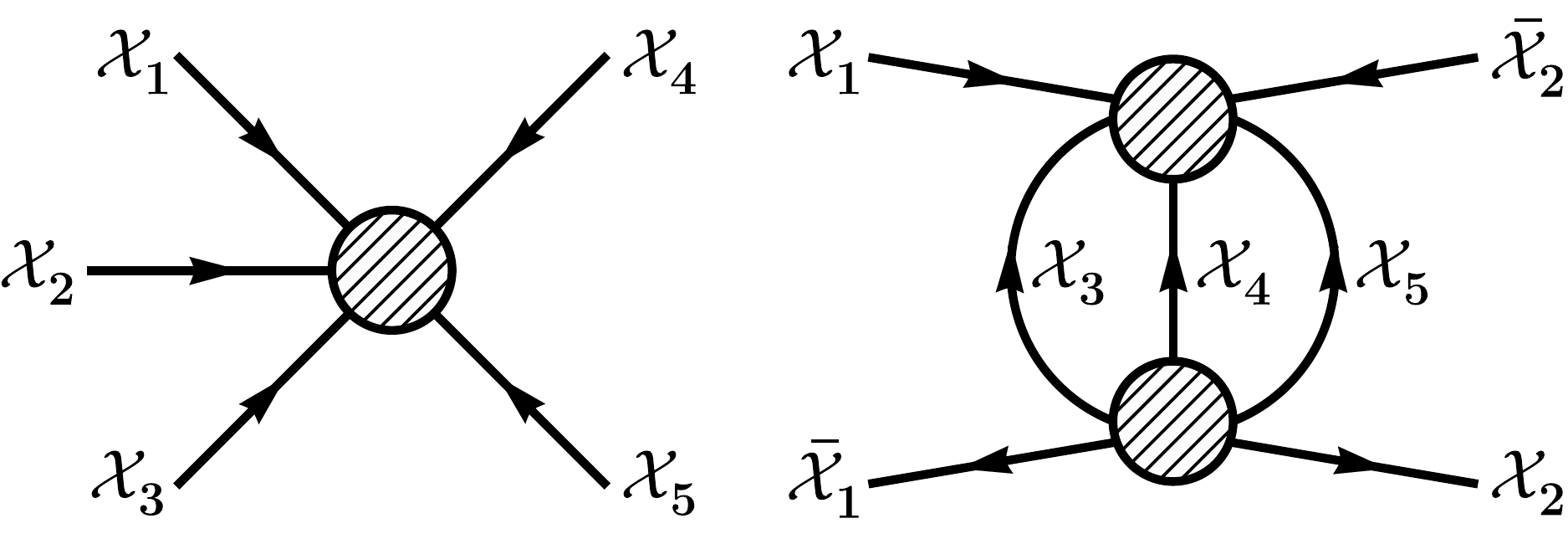}
\vs{-0.3cm}
\caption{The annihilation processes of multi-component SIMP DM ${\cal X}^{}_i$\,(which can be bosonic or fermionic) in the dark sector, where the arrows represent the direction of dark charge flow.\,\,The $3 \to 2$  processes (left graph) are induced by a five-point effective operator, which would inexorably generate the $2 \to 2$ processes through a two-loop topology (right graph).
\label{Fig:$r$SIMP}}
\end{center}
\vs{-0.5cm}
\end{figure}


Based upon the permutations of the interaction rates of DM and SM particles, the classification of thermal DM can be summarized as in Tab.\,\ref{tab:taxonomy}, where $H^{}_\tf{DM} \simeq m^2_\tf{DM}/m^{}_\tf{Pl}$ is the Hubble scale at which DM becomes nonrelativistic with $m^{}_\tf{DM}$ the DM mass and $m^{}_\tf{Pl} = 2.4\times 10^{18}\,\text{GeV}$ the reduced 
Planck mass, $\Gamma^{}_\tf{el} \equiv n^{}_\tf{SM}\langle \sigma^{}_\tf{el} v \rangle$ is the elastic scattering rate of DM with the SM particle, $\Gamma^{}_\tf{ann} \equiv n^{}_\tf{DM}\langle \sigma^{}_\tf{ann} v \rangle$ is the annihilation rate of a DM pair into SM particles, and $\Gamma^{}_{3 \to 2} \equiv n^2_\tf{DM} \langle \sigma^{}_{3 \to 2} v^2 \rangle$ is the $3 \to 2$ annihilation rate of DM.


The majority of DM models suggest that DM particle is WIMP-type and of only one kind.\,\,Nonetheless, the null results of direct search detections have cornered WIMP DM.\,${}^{}$Also, there is a possibility that the dark sector may be plentiful as same as the visible sector.\,\,Accordingly, it is reasonable to consider a scenario containing more than one species of DM  beyond the WIMP paradigm~\cite{Hochberg:2014kqa,Hochberg:2015vrg,Hochberg:2018rjs,Katz:2020ywn,Choi:2021yps,
Baek:2013dwa,Aoki:2016glu,Daido:2019tbm,Herms:2019mnu}.


The hidden quantum chromodynamics (HQCD) is an example possessing a lot of dark mesons that can serve as either WIMP \cite{Hur:2007uz,Ko:2008ug,Bai:2010qg,Hur:2011sv,Bai:2013xga,Hatanaka:2016rek} or SIMP DM particles \cite{Hochberg:2014kqa,Hochberg:2015vrg,Hochberg:2018rjs}.\,\,In HQCD models, the masses of dark mesons should be almost equal in order to give non-negligible contributions to the observed DM density.\,\,Otherwise, the heavier dark mesons would annihilate into the lighter ones via the inevitable $2 \to 2$ processes at leading order of chiral perturbation theory, leaving one-component SIMP DM~\cite{Katz:2020ywn}. 


In genuine multi-component WIMP or SIMP models, it should be possible for different DM species to have distinctive properties such as mass, (dark) charge, and even spin.\,\,In that sense, dark mesons in HQCD are not really multi-component DM models because they are related to each other by underlying hidden flavor symmetry.\,\,In the most recent paper of mSIMP model~\cite{Choi:2021yps}, the DM density is determined by the WIMP-like $2 \to 2$ and number-changing $3 \to 2$ processes with different species of DM to have a viable mSIMP scenario.


In this work, we propose a new type mSIMP scenario, $r$SIMP, where the relic density of DM is still determined by the $3 \to 2$ reaction rates, and we find that the degeneracy of SIMP masses is a necessary condition to have a $r$SIMP model.\,\,Namely, relative fine tuning of the mass spectra in the dark sector is required to make each DM component has a sizable amount of relic abundance.\,\,The key point is that even we can turn off all of $2 \to 2$ processes induced by four-point interactions and only retain $3 \to 2$ processes in the dark sector.\,\,The five-point interaction giving rise to the $3 \to 2$ processes can also generate number-conserving $2 \to 2$ processes at two-loop level as shown in Fig.\,\ref{Fig:$r$SIMP}.\,\,Moreover, the annihilation rates of the $2 \to 2$ processes would dominate over that of the $3 \to 2$ processes after the chemical freeze-out and redistribute the number densities of SIMP DM species. 


To see this feature, let us first parameterize the $3 \to 2$ annihilation rate as $\Gamma^{}_{3 \to 2} = n^2_\tf{DM} \alpha^3_\tf{eff}/m_\tf{DM}^5$, by which the annihilation rate of the two-loop induced $2 \to 2$ process can be approximated like $\Gamma^{2\tf{-loop}}_{2 \to 2} \simeq n^{}_\tf{DM}  \alpha^6_\tf{eff}/\big[(4\pi)^8 m_\tf{DM}^2\big]$, where $\alpha^{}_\tf{eff} \simeq 1-10$ is the effective strength of the five-point interactions, and $(4\pi)^8$ is the two-loop suppression factor.\,\,Making a comparison of these two reaction rates around the freeze-out temperature, $T^{}_f$, we find that
\begin{eqnarray}
\frac{\Gamma^{}_{3 \to 2}}{\Gamma^{2\tf{-loop}}_{2 \to 2}}
\Bigg|_{T{}^{}={}^{}T^{}_f}
=\,
\frac{(4\pi)^8}{\alpha^3_\tf{eff}}
\frac{n^{}_\tf{DM}(T^{}_f)}{m_\tf{DM}^3}
\,\simeq\,
\frac{1}{\alpha^3_\tf{eff} {}^{}{}^{} z_f^3}
\,\ll\, 1~,
\end{eqnarray}
where $n^{}_\tf{DM}(T) \simeq [{}^{}z/(2\pi)]^{3/2} {}^{} e^{-z} {}^{} T^3$, and $z^{}_f \equiv m^{}_\tf{DM}/T_f \simeq 20$.\,\,Having this estimate, we can classify the $r$SIMP as a thermal DM scenario with the condition below
\begin{eqnarray}
\Gamma^{2\tf{-loop}}_{2 \to 2} > \Gamma^{}_\tf{el} > \Gamma^{}_{3 \to 2} \gg \Gamma^{}_\tf{ann} > H^{}_\tf{DM}
~,
\end{eqnarray}
where the hierarchy of $\Gamma^{2\tf{-loop}}_{2 \to 2}$ and $\Gamma^{}_\tf{el}$ will be clear when we discuss the kinetic equilibrium between the dark and visible sectors.


For a systematic study of the $r$SIMP scenario, we will consider a few five-point effective operators consisting of scalar and fermions charged under a dark U$(1)^{}_\tf{D}$ symmetry, and solve the full Boltzmann equations including the $3 \to 2$ and two-loop induced $2 \to 2$ processes to obtain the correct cosmological evolutions of SIMP DM.\,\,Moreover, we find that the reshuffled mechanism in our fermion and scalar multi-component SIMP models can sensitively change the size of the DM self-interacting cross section which is not common in the single-component SIMP or HQCD models.
\vs{-0.1cm}


\section{Effective operators}
\vs{-0.1cm}
\begin{table}[t!] 
\vs{0.1cm}
\centering
\begin{tabular}{|c|c|c|c|}
\hline
\,Model\, & Fields & \,U$(1)^{}_\tf{D}$ & Interactions \\[0.05cm]
\hline
\multirow{1}*[-0.07cm]A & \multirow{1}*[-0.06cm]{$(X,N)$} & \multirow{1}*[-0.06cm]{$(2,-3)$} & 
\rule{0pt}{13pt} $\displaystyle{\cal O}^{(6)}_\tx{A}
=
\frac{c}{3! {}^{} 2! {}^{} \Lambda^2} 
X^3 \overline{N\raisebox{0.5pt}{$^\tf{c}$}} N$\,  \\[0.15cm]
\hline
\multirow{1}*[-0.07cm]B & \multirow{1}*[-0.06cm]{$(X,N^{}_1,N^{}_2)$} & \multirow{1}*[-0.06cm]{$(1,-2,-5)$} & 
\,\,\rule{0pt}{13pt}$\displaystyle{\cal O}^{(6)}_\tx{B}
=
\frac{c}{3! {}^{} \Lambda^2} 
X^3 \overline{\displaystyle N^{}_1} N^{}_2$\, \\[0.15cm]
\hline
\end{tabular}
\caption{Two representative models with relevant fields, U$(1)^{}_\tf{D}$ charges and interactions  in the dark sector, where ${\cal O}^{(6)}$ are dim-6 operators.\,\,Here we have omitted the hermitian-conjugate part for each ${\cal O}^{(6)}$.}
\label{tab:operator}
\vs{-0.0cm}
\end{table}

\renewcommand{\arraystretch}{1.45}
\begin{table}[t!] 
\centering
\begin{tabular}{|c|c|c|}
\hline
\,Model\, & $\,n \to 2\,$  & Number-conserving/changing process \\
\hline
\multirow{2}{*}[-0.04cm]A & $2 \to 2$ & 
$\XXNN, \NNXX$ 
\\ \cline{2-3}
                          & $3 \to 2$ & 
$\,\XXXNN, \XXNXN, \XNNXX$ 
\\\hline
\multirow{4}*[0.06cm]B & \multirow{2}*{$2 \to 2$} & 
$X \hs{-0.03cm} \bar{X} \to N^{}_{1,2} \bar{N}^{}_{1,2}{}^{}, N^{}_{1,2} \bar{N}^{}_{1,2} \to X \hs{-0.03cm} \bar{X}$
\\[-0.08cm]
                          &                                       & 
$N^{}_1 \bar{N}^{}_1 \to  N^{}_2 \bar{N}^{}_2{}^{},N^{}_2 \bar{N}^{}_2 \to  N^{}_1 \bar{N}^{}_1$ 
\\ \cline{2-3}
                          & \multirow{2}*{$3 \to 2$} &
$\,X \hs{-0.03cm} X \hs{-0.03cm} X \to \bar{N}^{}_2 N^{}_1{}^{}, 
X \hs{-0.03cm} X \hs{-0.03cm} N^{}_2 \to \bar{X} \hs{-0.03cm} N^{}_1$ 
\\[-0.08cm]
                          &                                       &
$X \hs{-0.03cm} X \hs{-0.03cm} \bar{N}^{}_1 \to \bar{X} \hs{-0.03cm} \bar{N}^{}_2{}^{},
X \hs{-0.03cm} N^{}_2 \bar{N}^{}_1 \to \bar{X} \hs{-0.03cm} \bar{X}$
\\\hline
\end{tabular}
\caption{All kinematically allowed annihilation processes in two benchmark models, here we have omitted the charge-conjugation processes in this table.}
\label{tab:ncp}
\vs{-0.2cm}
\end{table}


In our setup, instead of enumerating all possible symmetries and interactions, here we consider two representative five-point effective operators which can generate $3 \to 2$ annihilations of different DM species in the dark sector.\,\,We introduce one complex scalar $X$, and Dirac fermions $N,N^{}_1,N^{}_2$ with proper dark U$(1)^{}_\tf{D}$ charges as SIMP DM candidates.\,\,Our two benchmark models are shown in Tab.\,\ref{tab:operator}, where Model A is for a two-component DM scenario and Model B is for a three-component DM scenario.\,\,Note that, with these dark charge assignments, there is no extra five-point interaction and no mass mixing for each model.


In these five-point operators, the $c$ is a dimensionless constant and the $\Lambda$ is a cut-off  energy scale.\,\,Here we presume that both ${\cal O}^{(6)}_\tx{A}$ and ${\cal O}^{(6)}_\tx{B}$ effective operators in Tab.\,\ref{tab:operator} come from some UV complete models after integrating out the heavy mediators.\,\,Thus, we expect that the $\Lambda$ is associated with mediator masses and the $c$ is related to coupling products in UV complete models.\,\,We will show a concrete and simple toy model to check this point in the latter section.\,\,Note that we ignore the four-point interactions such as $|X|^2 {}^{} \overline{N} N$ since they can be switched off without affecting the five-point interactions.


With these interactions, we display all possible $2 \to 2$ and $3 \to 2$ annihilation processes for Models A and B in Tab.\,\ref{tab:ncp}, where the mass relations $3{}^{}m^{}_X \hs{-0.08cm} >  2{}^{}m^{}_{N} \hs{-0.08cm} > m^{}_X$ for Model A, and $3{}^{}m^{}_X > m^{}_{N_1} \hs{-0.05cm} + m^{}_{N_2} > m^{}_X > \big|m^{}_{N_1} \hs{-0.05cm} - m^{}_{N_2}\big|$ for Model B are imposed, under which the $2 \to 3$ and $2 \to 4$ processes such as $\NNXXX$ and $\XXNNNN$, etc. are kinematically forbidden.

 
Also, if this dark U$(1)^{}_\tf{D}$ symmetry is promoted to be gauged, then it is natural to have a dark gauge boson which induces vector portal effective operators between the DM and SM fermions $f$ after the U$(1)^{}_\tf{D}$ symmetry breaking as~\cite{Lehmann:2020lcv}
\begin{eqnarray} 
{\cal O}^{(6)}_{\hs{-0.03cm} X\hs{-0.03cm} f}
\Eq
\frac{i{}^{}c^{}_{X\hs{-0.03cm} f}}{\Lambda^2_{Z'}} 
\big(
X^{\dagger}\partial_\mu X - X \partial_\mu X^{\dagger}
\big) 
\big(
\overline{f} {}^{} \gamma^\mu f
\big)
~,
\label{cphie}
\\[0.1cm]
{\cal O}^{(6)}_{\hs{-0.03cm} \psi f}
\Eq 
\frac{c^{}_{\psi f}}{\Lambda^2_{Z'}}
\big({}^{}{}^{}
\overline{\psi} {}^{} \gamma^{}_\mu \psi
\big) 
\big(
\overline{f} {}^{} \gamma^\mu f
\big)
~,
\quad \psi \,=\, N, N^{}_1, N^{}_2
~, 
\label{cpsie}
\end{eqnarray}
where $c^{}_{X\hs{-0.03cm}f,{}^{}\psi f}$ are dimensionless couplings, and $\Lambda^{}_{Z'}$ characterizes the mass scale of the vector mediator.\,\,Because of the DM mass scale in this scenario, here we focus on the DM and $f = e^{\pm}$ interactions written above.\,\,Notice that with these electrophilic interactions, the SIMP DM masses $\lesssim {\cal O}(10)\,\tx{MeV}$ are disfavored due to the cosmological observation for the effective number of neutrino species \cite{Smirnov:2020zwf}.
\vs{-0.2cm}


\section{Cosmological evolution}

In this section, we will write down the full Boltzmann equations of DM comoving number densities and explain the behaviors of their evolutions in both models.

Based upon the annihilation processes as displayed in Tab.\,\ref{tab:ncp}, we derive the 
Boltzmann equations of comoving number yields $Y^{}_X$ and $Y^{}_N$ for $X$ and $N$, respectively, in Model A as functions of the dimensionless time variable, $x \equiv m^{}_X/T$, as below (assuming $Y^{}_{X,N} = Y_{\bar{X},\bar{N}}$)
\begin{eqnarray}
\frac{\dd Y^{}_X}{\dd x}
\Eq
-\frac{s^2}{H x}
\Bigg\{
12 \langle \XXXNN \rangle \hs{-0.03cm}
\sx{1.1}{\bigg[} 
Y^3_X - Y_N^2 \frac{(Y^\textsf{eq}_X)^3}{(Y^\tf{eq}_N)^2} 
\sx{1.1}{\bigg]}
\nn
&&\hs{1.2cm}
{+}\,2\langle \XXNXN \rangle 
\sx{1.1}{\big(} 
Y^2_X Y^{\white{2}}_N - Y^{}_X Y^{\white{2}}_N Y^\tf{eq}_X
\sx{1.1}{\big)}
\nn
&&\hs{1.2cm}
{-}\,\langle\XNNXX \rangle \hs{-0.03cm}
\sx{1.1}{\bigg[} 
Y^{}_X Y_N^2 - Y^2_X \frac{(Y^\tf{eq}_N)^2}{Y^\tf{eq}_X} 
\sx{1.1}{\bigg]} \hs{-0.05cm}
\Bigg\}
\nn
&&
-\frac{s}{H x}
\Bigg\{
4\langle \XXNN \rangle \hs{-0.03cm}
\sx{1.1}{\bigg[} 
Y_X^2 - Y_N^2 \frac{(Y^\tf{eq}_X)^2}{(Y^\tf{eq}_N)^2} 
\sx{1.1}{\bigg]} 
\nn
&&\hs{1.2cm}
{-}\,\langle \NNXX \rangle
\sx{1.1}{\bigg[} 
Y_N^2 - Y_X^2 \frac{(Y^\tf{eq}_N)^2}{(Y^\tf{eq}_X)^2} 
\sx{1.1}{\bigg]} \hs{-0.05cm}
\Bigg\}
~,
\label{dYX}
\\[0.2cm]
\frac{\dd Y^{}_N}{\dd x} 
\Eq
-\frac{s^2}{H x}
\Bigg\{
2\langle \XNNXX \rangle \hs{-0.03cm}
\sx{1.1}{\bigg[} 
Y^{}_X Y^2_N - Y^2_X \frac{(Y^\tf{eq}_N)^2}{Y^\tf{eq}_X} 
\sx{1.1}{\bigg]}
\nn
&&\hs{1.2cm}
{-}\,8\langle \XXXNN \rangle \hs{-0.03cm}
\sx{1.1}{\bigg[} 
Y_X^3 - Y_N^2 \frac{(Y^\tf{eq}_X)^3}{(Y^\tf{eq}_N)^2} 
\sx{1.1}{\bigg]} \hs{-0.05cm}
\Bigg\}
\nn
&&
-\frac{s}{H x}
\Bigg\{
\langle \NNXX \rangle \hs{-0.03cm}
\sx{1.1}{\bigg[} 
Y_N^2 - Y_X^2 \frac{(Y^\tf{eq}_N)^2}{(Y^\tf{eq}_X)^2} 
\sx{1.1}{\bigg]} 
\nn
&&\hs{1.2cm}
{-}\,4\langle \XXNN \rangle \hs{-0.03cm}
\sx{1.1}{\bigg[} 
Y_X^2 - Y_N^2 \frac{(Y^\tf{eq}_X)^2}{(Y^\tf{eq}_N)^2} 
\sx{1.1}{\bigg]} \hs{-0.05cm}
\Bigg\}
\,,
\label{dYN}
\end{eqnarray}
where $\langle ijk \hs{-0.08cm} \to \hs{-0.08cm} lm \rangle = \langle \sigma \upsilon^2 \rangle^{}_{ijk \to lm}$, $\langle ij \hs{-0.08cm} \to \hs{-0.08cm} lm \rangle = \langle \sigma \upsilon \rangle^{}_{ij \to lm}$ are thermally-averaged cross sections, $Y^\tf{eq}_i$ is the equilibrium comoving number density of the DM species $i$ with the internal degrees of freedom $g^{}_i$,
\begin{eqnarray}
\hs{-0.5cm}
Y^\tf{eq}_i
\approx
\frac{45\sqrt{2}}{8{}^{}\pi^{7/2}}
\frac{g^{}_i}{g^{}_{\star s}(x)} {}^{}
(r^{}_i {}^{} x)^{3/2} e^{- r^{}_i {}^{} x}
~,\quad 
r^{}_i \equiv \frac{m^{}_i}{m^{}_X}
\end{eqnarray}
with $g^{}_{\star s}(x)$ being the entropy degrees of freedom of the thermal plasma~\cite{Saikawa:2018rcs}, $s$ is the comoving entropy density, and $H$ is the Hubble parameter\,\cite{Kolb:1990vq}.


As we shall see soon, the SIMP DM masses should be degenerate to contribute a non-negligible amount to the total DM relic density.\,\,Given the five-point interactions in Tab.\,\ref{tab:operator}, the $3 \to 2$ annihilation cross sections with the degenerate DM masses are calculated as\,\cite{Yang:2019bvg}
\begin{eqnarray}
\langle \XXXNN \rangle
\Eq
\frac{\sqrt{5} {}^{}{}^{} c^2 (m^{}_X/\Lambda)^4}{4608{}^{}\pi{}^{}m_X^5} 
\bigg( \hs{-0.03cm} 5 + \frac{18}{x} + \frac{12}{x^2} \bigg)
~,\quad
\\[0.1cm]
\langle \XXNXN \rangle
\Eq
\frac{\sqrt{5} {}^{}{}^{} c^2 (m^{}_X/\Lambda)^4}{768{}^{}\pi{}^{}m_X^5} 
\bigg( \hs{-0.03cm} 5 + \frac{6}{x} + \frac{4}{x^2} \hs{-0.04cm} \bigg)
~,
\\[0.1cm]
\langle \XNNXX \rangle
\Eq
\frac{\sqrt{5} {}^{}{}^{} c^2 (m^{}_X/\Lambda)^4 }{768{}^{}\pi{}^{}m_X^5} 
\bigg( {}^{} \frac{3}{x} + \frac{2}{x^2} \hs{-0.05cm} \bigg)
~,
\end{eqnarray}
here we have used the SO(9) invariant form of the total kinetic energy in the initial states~\cite{Choi:2017mkk} and the Feynman rules of fermion-number-violating interactions in~\cite{Denner:1992vza} for deriving these cross sections.\,\,On the other hand, using the integration result of the two-loop sunset graph with $\Lambda \gtrsim m^{}_{X,N}$\,\cite{Yang:2003bv}, the $2 \to 2$ annihilation cross sections with the degenerate DM masses are computed as
\begin{eqnarray}
&&\hs{-0.7cm}
\langle \XXNN \rangle
=
\frac{c^4 (m^{}_X/\Lambda)^4}{16 {}^{} \pi {}^{} (4\pi)^8 m_X^2} {}^{}{}^{}
{\cal I}^{{}^{}2}_\Lambda \hs{-0.05cm} 
\sx{0.9}{\bigg(} \frac{\Lambda}{m^{}_X} \hs{-0.02cm} \sx{0.9}{\bigg)}
\mathord{\raisebox{0.5\depth}{$\sqrt{1-r_N^2}$}}
\nn
&&\hs{1.6cm}
\times \hs{-0.1cm} \bigg[1-r_N^2 + \frac{3}{4 {}^{} x} \big( 5{}^{}r_N^2-2 \big) \bigg]
~,
\\[0.1cm]
&&\hs{-0.7cm}
\langle \NNXX \rangle
=
\frac{3 {}^{} c^4 (m^{}_X/\Lambda)^4}{32 {}^{} \pi {}^{} (4\pi)^8 x {}^{}{}^{} m_X^2} {}^{}
{\cal I}^{{}^{}2}_\Lambda \hs{-0.05cm} 
\sx{0.9}{\bigg(} \frac{\Lambda}{m^{}_X} \hs{-0.02cm} \sx{0.9}{\bigg)}
\frac{\sqrt{r^2_N-1}}{r^{}_N} 
~,
\end{eqnarray}
where the two-loop function with ${\cal I}^{}_\Lambda(\infty) = 1$ reads\,\cite{Yang:2003bv}
\begin{eqnarray}
{\cal I}^{}_\Lambda(t)
\,\approx\,
1 + \frac{1}{2{}^{}t^2} 
\bigg[
4 - 3 \ln t^2 - \frac{3}{2} \big( \hs{-0.03cm} \ln t^2 \big)^{\hs{-0.05cm}2}
\bigg]
~.
\end{eqnarray}
Note that the $\langle \XNNXX \rangle$ and $\langle \NNXX \rangle$ are dominated by $p\,$-wave contributions.

\begin{figure}[t!]
\begin{center}
\includegraphics[width=0.48\textwidth]{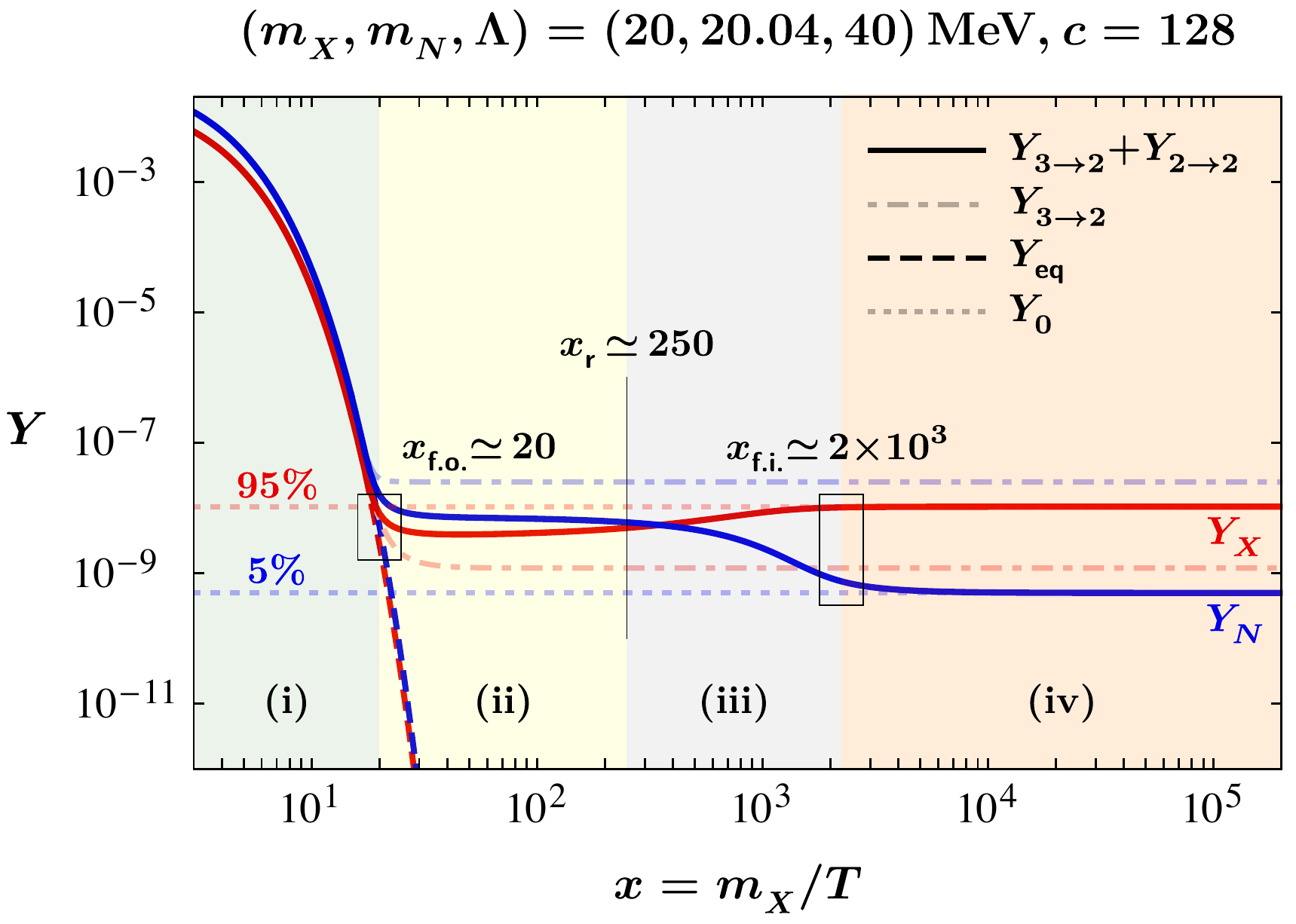}
\\[0.3cm]
\includegraphics[width=0.48\textwidth]{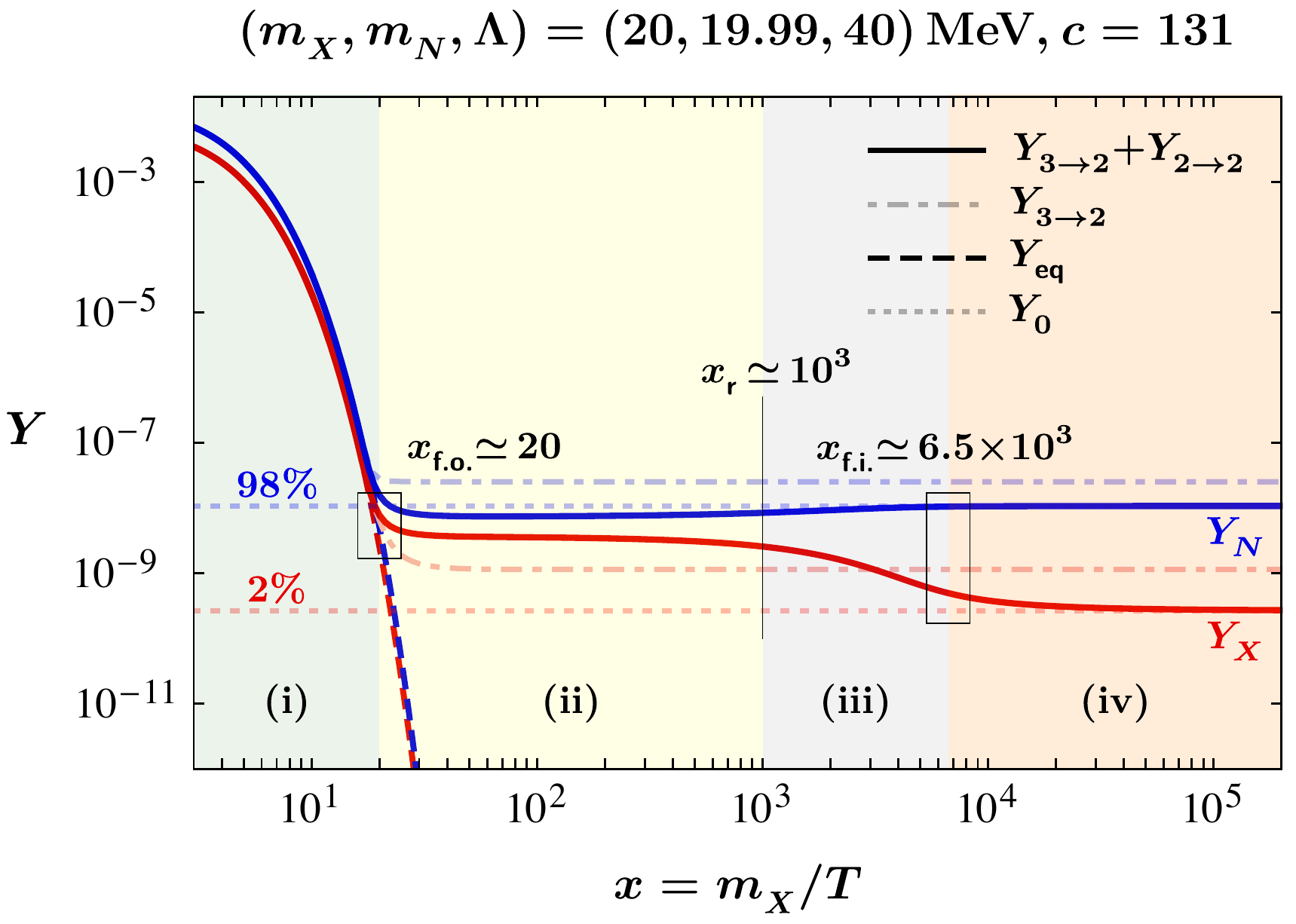}
\vs{-0.5cm}
\caption{The cosmological evolution of the comoving number densities of $X$ and $N$ in Model A with two benchmark cases, where the solid (dot-dashed) line is the DM number yield in the case with (without) $2 \to 2$ reactions, the dashed (dotted) line denotes the DM number density at chemical equilibrium (present), $x^{}_\tf{r}$ is the reshuffle temperature, and $\xfo\,(\xfi)$ is the freeze-out (-in) temperature of SIMP DM particles.}
\label{fig:YXN}
\end{center}
\vs{-0.3cm}
\end{figure}

Next, with a proper initial condition of the Boltzmann equations, we can obtain the $Y^{}_{X,N}(x)$, and then predict the current density of DM by the following relation\,\cite{Bhattacharya:2019mmy} 
\begin{eqnarray}
\Omega^{}_\tf{DM} h^2 
\,\simeq\,
5.49 \times 10^5 {}^{}
\sx{0.9}{\bigg(}
\frac{m^{}_X}{\text{MeV}}
\sx{0.9}{\bigg)}
\sum_{i{}^{}={}^{}X,N}
Y^0_i {}^{} r^{}_i 
~,
\end{eqnarray}
where $Y^0_i = Y^{}_i({}^{} x\to\infty)$ denotes the present value of $Y^{}_i$.\newline 
We present the numerical solutions for Eqs.\,\eqref{dYX} and \eqref{dYN} with two benchmark 
points in Fig.\,\ref{fig:YXN}, where the color solid lines satisfy the observed relic density of DM, $\Omega^\tf{obs}_\tf{DM} h^2 = 0.12 \pm 0.0012$ \cite{Aghanim:2018eyx}.\,\,As expected, 
the $X$ and $N$ can have sizable contributions if their masses are degenerate.

The evolution of the comoving number density can be divided into four stages as shown in color shaded regions of Fig.\,\ref{fig:YXN}.\,\,In region (i), the effective $3 \to 2$ reaction rate is much larger than the Hubble expansion rate, $\Gamma^{}_{3 \to 2} \gg H$, where the DM number densities are depleted due to the $3 \to 2$ processes until the freeze-out temperature, $\xfo \hs{-0.08cm} \simeq 20$, at which $\Gamma^{}_{3 \to 2} \simeq H$.\,\,In region (ii), the DM particles depart from the chemical equilibrium since $\Gamma^{}_{3 \to 2} \ll H$. Now, as we mentioned earlier, the $\Gamma^{2\tf{-loop}}_{2 \to 2}$ dominates over $\Gamma^{}_{3 \to 2}$ in most of the time of the universe.\,\,However, due to the degeneracy of DM masses in the $r$SIMP scenario, the forward $2 \to 2$ reaction is cancelled by the backward one.\,\,This is the reason that the number densities of DM maintain constants for a while after the chemical freeze-out.\,\,To illustrate this point explicitly, let us look at the last term of \eqref{dYX},\,$\langle \NNXX \rangle \big[Y^2_N - 4Y^2_X r_N^3 {}^{} e^{-2(r^{}_N-1)x}\big]$, where the first (second) term in the square bracket is the rate for the forward (backward) $2 \to 2$ reaction.\,\,Clearly, at high temperatures, $r_N^3 {}^{} e^{-2(r^{}_N-1)x} \sim 1$ if $r^{}_N$ is close to 1, while $Y^{}_N \hs{-0.05cm} \sim 2Y^{}_X$ right after the freeze-out.\,\,As a result, this term vanishes and gives no physical effect until $x \simeq x^{}_\tf{r} \hs{-0.03cm} \equiv 1/(2|r^{}_N-1|)$, after that the backward reaction rate is exponentially-suppressed.\,\,In region (iii), the forward $2 \to 2$ reaction becomes active, the heavier DM particles annihilate into the lighter ones during this period.\,\,Note that since the $2 \to 2$ processes preserve the total DM number density, and so, it would merely redistribute the number densities of DM until the freeze-in temperature, $\xfi$, where $\Gamma^{2\tf{-loop}}_{2 \to 2} \simeq H$.\,\,In region (iv), the DM number densities are frozen until today.

\begin{figure}[t]
\begin{center}
\includegraphics[width=0.48\textwidth]{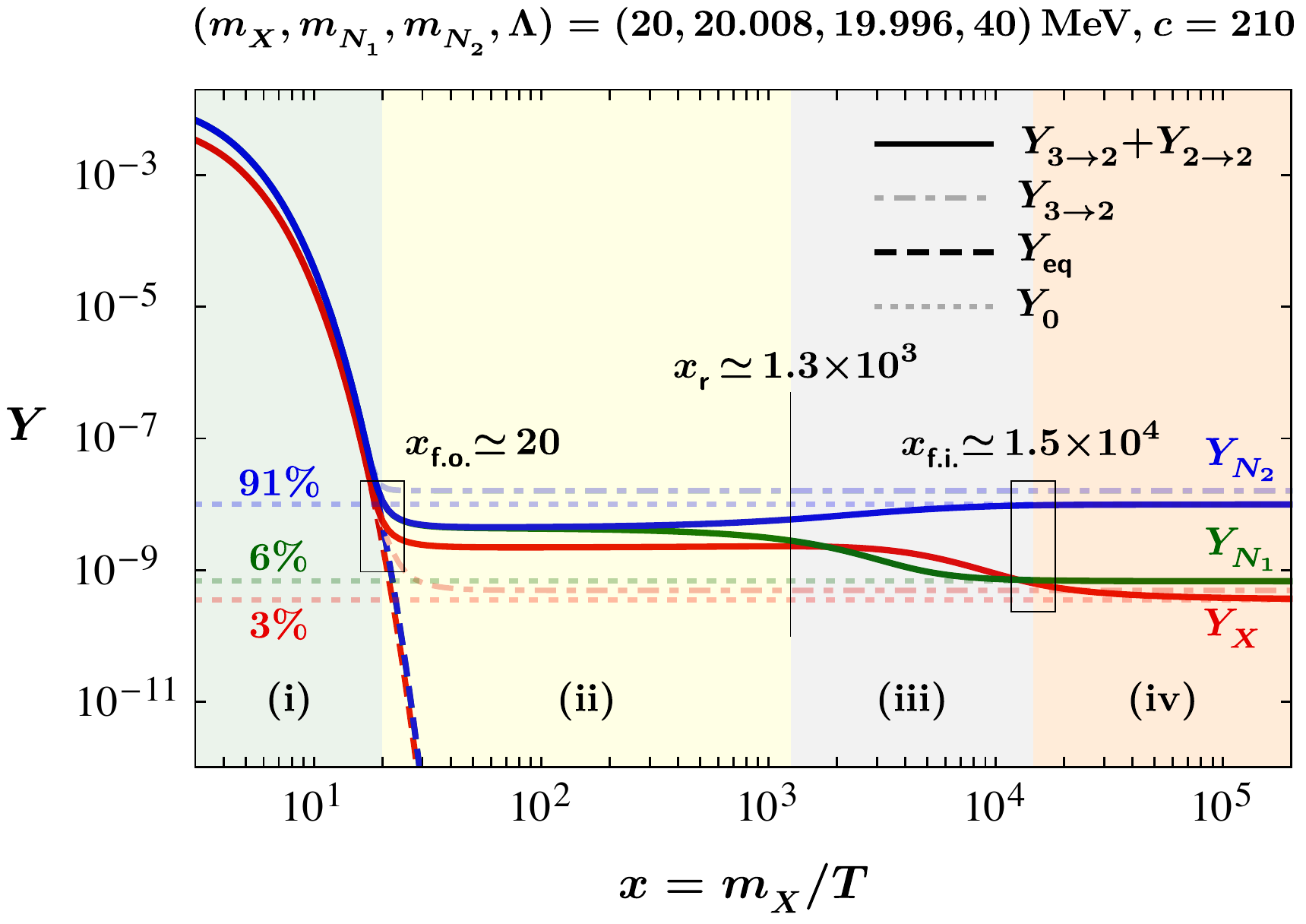}
\\[0.3cm]
\includegraphics[width=0.48\textwidth]{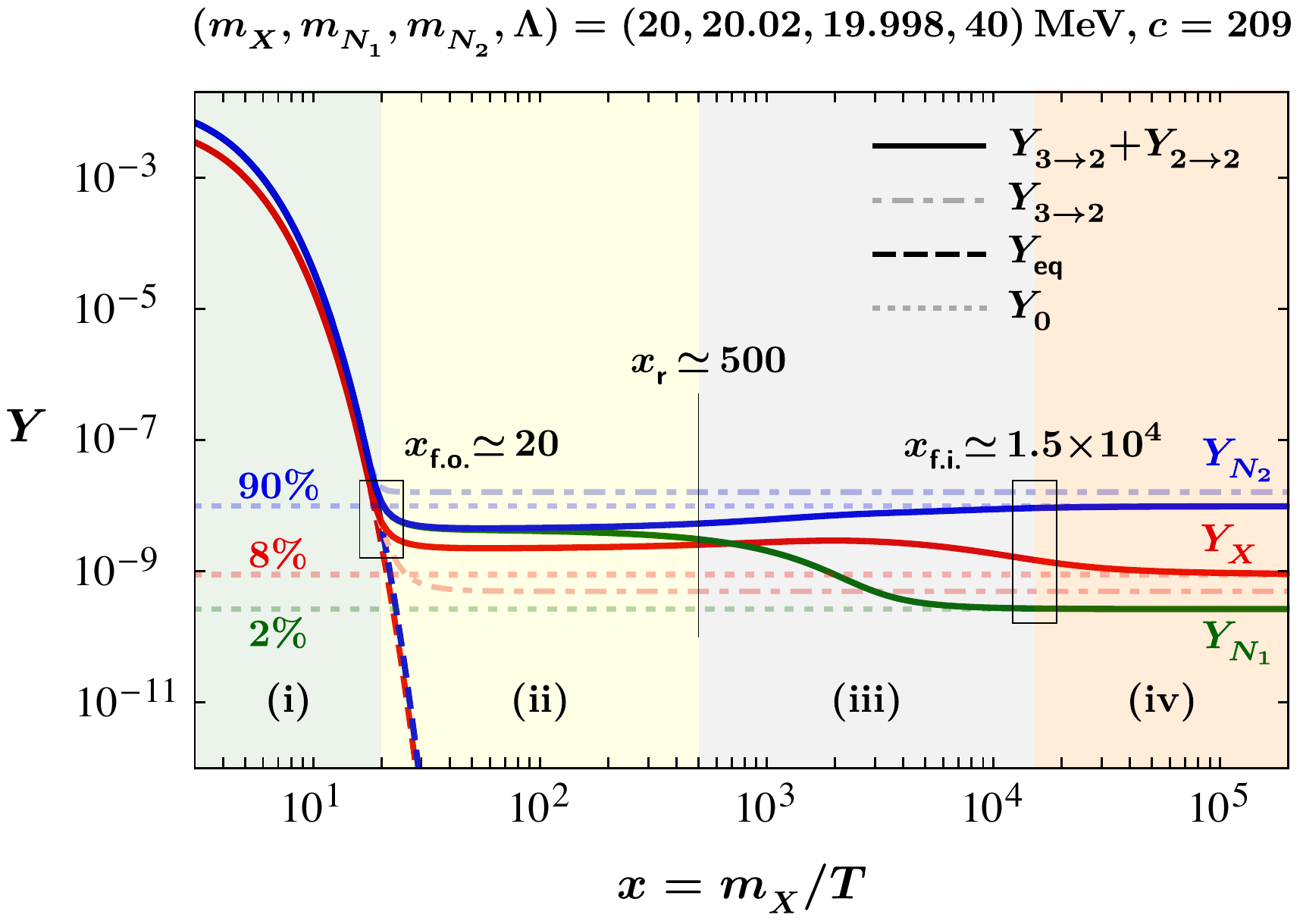}
\\[0.3cm]
\includegraphics[width=0.48\textwidth]{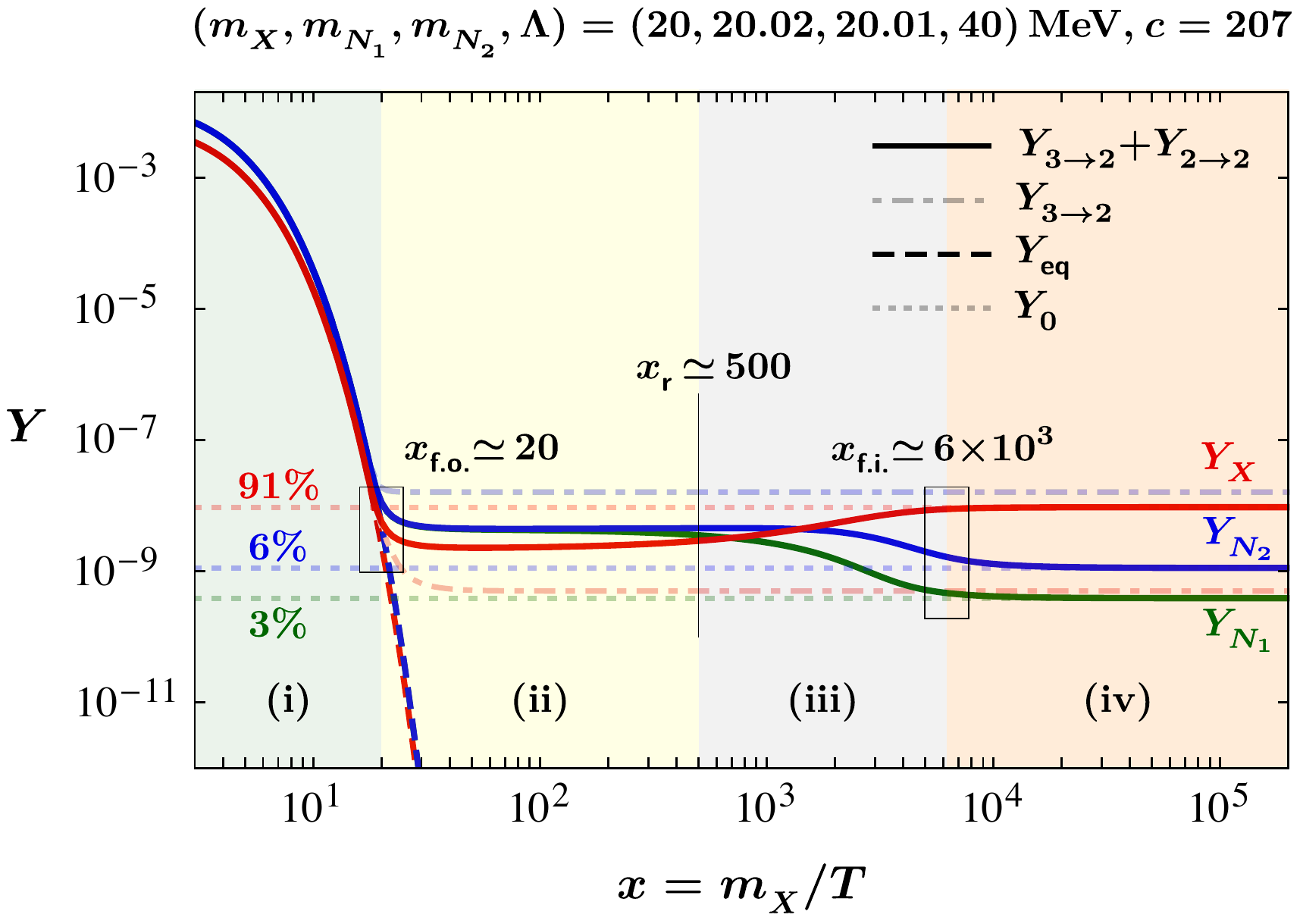}
\vs{-0.5cm}
\caption{The cosmological evolution of the comoving number densities in Model B.\,\,Upper panel\,\,\,:\,\,$r^{}_N < 1$.\,\,Middel panel\,\,:\,\,$r^{}_1 > 1 > r^{}_2$.\,\,Bottom panel\,\,:\,\,$r^{}_1 > r^{}_2 > 1$.}
\label{fig:YXN2N1}
\end{center}
\vs{-0.5cm}
\end{figure}

Likewise, we also derive the Boltzmann equations describing the comoving number densities $Y^{}_X$ and $Y^{}_{N_{1,2}}$ for $X$ and $N^{}_{1,2}$, respectively, in Model B.\,\,However, since the Boltzmann equations and the cross sections are lengthy, we collect them into the appendix.\,\,We show in Fig.\,\ref{fig:YXN2N1} a few benchmark examples of the evolution of the comoving number densities in Model B.\,\,As can be seen from these plots, the reshuffled phenomena of the DM number densities is very remarkable.

The thermally-averaged $3 \to 2$ cross sections can be bounded from above by partial-wave unitarity.\,\,Quoting the result in \cite{Namjoo:2018oyn}, we obtain the strictest bounds, where $c\,(m^{}_X/\Lambda)^2 \lesssim 8 \times 10^3$ for both models.\,\,Hence, our benchmark points in Figs.\,\ref{fig:YXN} and \ref{fig:YXN2N1} are justified.\,\,Of course, one can consider heavier DM masses with a much larger $c$ to satisfy the relic abundance of DM and unitarity constraint.\,\,However, the couplings corresponding to $c$ may be subject to perturbative limits in UV complete models.\,\,Thus, we suggest that the typical DM masses in the $r$SIMP scenario are around ${\cal O}(20)\,\tx{MeV}$.

\section{Kinetic equilibrium}

As in the conventional SIMP scenario, the DM should keep kinetic equilibrium with the SM particles until the freeze-out temperature of DM, where the energy transfer rate $\gamma(x)$ of these two sectors fulfills the thermalization condition,\,\,$\gamma(z^{}_f) \gtrsim H(z^{}_f) {}^{} z_f^2$\,\cite{Choi:2019zeb}.\,\,From Eqs.\,\eqref{cphie} and \eqref{cpsie}, the energy transfer rate between the SIMP DM particles and $e^\pm$ with $r^{}_i \sim1$ is computed as~\cite{Gondolo:2012vh}
\begin{eqnarray}\label{gammae}
\gamma^{}_e(x)
=
\frac{31\pi^3}{189{}^{}{}^{}x^6}
\frac{m_X^5}{\Lambda^4_{Z'}}
\sum_{i\,=\,\tf{SIMP}} c_{ie}^2
~.
\end{eqnarray}
Using the condition of kinetic equilibrium, we arrive at
\begin{eqnarray}\label{cebound}
\sum_{i\,=\,\tf{SIMP}} c_{ie}^2
\gtrsim
10^{-9}
\bigg(\frac{ \Lambda^{}_{Z'}}{200\,\text{MeV}}\bigg)^{\hs{-0.13cm}4}
\bigg(\frac{m^{}_X}{20\,\text{MeV}}\bigg)^{\hs{-0.15cm}-3}
~.
\end{eqnarray}

In addition, the kinetic decoupling temperature of the dark sector from the visible sector can be determined by the relation $\gamma(\xkd) \simeq 2 H(\xkd)$\,\cite{Gondolo:2012vh}.\,\,Using the marginal value of the thermalization condition as well as Eq.\,\eqref{gammae}, we find  that the highest kinetic decoupling temperature, $\xkd \hs{-0.05cm} \simeq 75 \ll \xfi$, indicating that $\Gamma^{2\tf{-loop}}_{2 \to 2} > \Gamma^{}_\tf{el}$.


Now, let us comment on the gauge coupling $\gD$ of the U(1)$^{}_\tf{D}$ symmetry in these two models as it is related to the $c^{}_{ie}$ in Eq.\,\eqref{cebound}.\,\,If all the SM fields are neutral under the dark U(1)$^{}_\tf{D}$ symmetry, then $e^\pm$ can only couple to the new dark boson via the kinetic mixing between the U(1)$^{}_\tf{D}$ and the SM hypercharge U(1)$^{}_\tf{Y}$ gauge bosons.\,\,In this manner, one can define $c^{}_{ie} \hs{-0.03cm} \equiv -{}^{}\epsilon {}^{} g^{}_e {}^{} g^{}_\tf{D} Q^{}_{i,\tf{D}}$, where $\epsilon$ is the kinetic mixing parameter, $g^{}_e \sim 0.3$ is the electric charge, and $Q_{i,\tf{D}}$ is the U(1)$^{}_\tf{D}$ charge of the DM species $i$ assigned in Tab.\,\ref{tab:operator}.\,\,Based on Eq.\,\eqref{cebound} with the fiducial values $\Lambda^{}_{Z'} = 200$\,MeV and $m^{}_X = 20$\,MeV, we find that $\gD \gtrsim 0.03$ for Model A, and $\gD \gtrsim 0.02$ for Model B if $\epsilon \lesssim 10^{-3}$\,\cite{Fabbrichesi:2020wbt}.\,\,Besides, we have checked that there is no allowed parameter space for gauge couplings if the new U(1) gauge symmetry is U(1)$^{}_{\tf{B}-\tf{L}}$, U(1)$^{}_{\tf{L}_{\mu}-\tf{L}_{\tau}}$, etc\,\cite{Bauer:2018egk}. 

In fact, the SIMP particles can also annihilate among each other through the vector portal interactions akin to Eqs.\,\eqref{cphie} and \eqref{cpsie}.\,\,However, we find that the contribution of the vector portal diagram for the $2 \to 2$ process is subdominant to that of the two-loop diagram if we take the minimal values of $\,\gD\,$ found above.\,\,Note that we have to choose 
the $\gD$ value as small as possible to suppress the freeze-out processes, $X\hs{-0.03cm}\bar{X}, N\hs{-0.03cm}\bar{N} \to e^+e^-$, etc. of the WIMP scenario.\,\,Thus, the main source of the reshuffled effect comes from the two-loop induced diagrams in the $r$SIMP scenario.
\vs{-0.3cm}

\section{A toy model}
\vs{-0.1cm}
To realize the $r$SIMP scenario, we consider the following UV complete model for Model A as
\begin{eqnarray}
\hs{-0.3cm}
{\cal L}^{}_\tx{Model A}
\,=\,
-\tfrac{1}{6} \lambda^{}_3 X^3 \hs{-0.03cm} S^\ast 
-\tfrac{1}{2}{}^{}y^{}_N \overline{N\raisebox{0.5pt}{$^\tf{c}$}} \hs{-0.03cm} N S 
+\tx{h.c.}
~,
\end{eqnarray}
where $\lambda^{}_3 > 0$ is the quartic coupling, $y^{}_N > 0$ is the Yukawa coupling, and the U$(1)^{}_\tf{D}$ charge of the additional scalar particle $S$ is assigned to $6$.\,\,After integrating out the mediator $S$, we can generate the ${\cal O}^{(6)}_\tx{A}$ given in Tab.\,\ref{tab:operator}.\,\,Here we have to assume that $m^{}_S > 2{}^{}m^{}_N$ or $m^{}_S > 3{}^{}m^{}_X$.\,\,Otherwise, the $S$ particle may also be a DM candidate.

With this setup, we can explicitly compute the $3 \to 2$ and $2 \to 2$ annihilation cross sections, then numerically solve the Boltzmann equations to yield the correct DM number densities.\,\,In particular, we have confirmed that the form of the $2 \to 2$ annihilation cross section derived by the effective operator approach is consistent with the one in this UV complete model if we identify $c{}^{}\sim \lambda^{}_3{}^{}y^{}_N$ and $\Lambda \sim m^{}_S$.\,\,Thus, our previous argument for the $r$SIMP scenario by the effective theories is robust and reliable.\,\,Also, the size of $c$ can be reduced by the resonance effect \cite{Choi:2016hid}, say $m^{}_S \simeq 3{}^{}m^{}_X$, such that the perturbative bounds can be evaded in the UV complete model, see Ref.\,\cite{Ho:2022erb} for detail computations.

\section{The astrophysical signature}

In the SIMP DM models, a sufficiently large coupling is needed to satisfy the DM relic abundance.\,\,With such couplings, the predictions for DM self-interacting cross sections may be too large to be compatible with the astrophysical observations from the Bullet and Abell 3827 clusters \cite{Markevitch:2003at,Clowe:2003tk,Massey:2015dkw,Kahlhoefer:2015vua}.\,\,In the $r$SIMP scenario, this tension can be alleviated thanks to the reshuffled mechanism.\,\,For instance, the DM self-interacting cross section in the toy Model A is given by
\begin{eqnarray}
\frac{\sigma^{}_\tf{self}}{m^{}_\tf{DM}}
\,=\, 
{\cal R}^{}_X \frac{\sigma^{}_X}{m^{}_X} + 
{\cal R}^{}_N \frac{\sigma^{}_N}{m^{}_N} ~,
\end{eqnarray}
where ${\cal R}^{}_{X,N} \,=\, \Omega^{}_{X,N}/(\Omega^{}_X + \Omega^{}_N)$ is the fraction of each DM component, and 
\begin{eqnarray}
\sigma^{}_X \,=\, \frac{\lambda_X^2}{8\pi m_X^2} ~,\quad
\sigma^{}_N \,=\, \frac{y_N^4}{16\pi m_N^2}\bigg(\frac{m^{}_N}{m^{}_S}\bigg)^{\hs{-0.1cm}4} ~,
\end{eqnarray}
with $\lambda^{}_X$ the quartic couping of $X$.\,\,As defined, the DM self-interacting cross section is sensitive to the ratios of DM.\,\,Hence, if $m^{}_X > m^{}_N$, the DM self-interacting cross section can be reduced since the portion of the complex scalars annihilate into the vector-like fermions through the $2 \to 2 $ process, $\XXNN$ (see the bottom panel of Fig.\,\ref{fig:YXN}), while the self-interacting cross section of the vector-like fermion is suppressed by the mass of the mediator $S$.\,\,This is one of the appealing features of this scenario, see Ref.\,\cite{Ho:2022erb} for more discussions.\,\,Lastly, we want to point out that such a two-loop induced $2 \to 2$ process may affect some other thermal DM scenarios such as the Co-SIMP scenario \cite{Ho:2021pqw}.

\section{Conclusions}

In this paper, we propose a brand-new class of multicomponent thermally-produced DM, $r$SIMP, where the number densities of DM are redistributed after the DM chemical freeze-out.\,\,This is due to the two-loop induced number-conserving $2 \to 2$ processes inevitably generated by the five-point interactions for the $3 \to 2$ processes.\,\,We have studied this scenario by the effective operator approach and numerically solved the Boltzmann equations involving these processes to obtain the correct yields of SIMP.\,\,We find the DM masses must be nearly degenerate and around ${\cal O}(20\,\tx{MeV})$ in order to contribute non-negligible densities to the observed DM relic abundance.\,\,Remarkably, the prediction of the self-interacting cross section of DM is modified by the reshuffled effect, which can be tested by future astronomical observations and simulations.

\newpage
\acknowledgments
\section{Acknowledgments}
SYH would like to thank Hiroyuki Ishida for discussion. This work is supported by KIAS Individual Grants under Grant No.\,PG081201 (SYH), No.\,PG075301 (CTL), and No.\,PG021403 (PK), and also in part by National Research Foundation of Korea (NRF) Grant No.\,NRF-2019R1A2C3005009 (PK).

\newpage
\section{Appendix}

\subsection{Boltzmann equations}

In this appendix, we collect the full Boltzmann equations with the complete $3 \to 2$ and $2\to 2$ annihilation cross sections for Model B.

Schematically, the Boltzmann equation governing the number density 
of the DM species, $i$, including the $3 \to 2 $ and $2 \to 2$ annihilations is given by
\begin{eqnarray}\label{nBE}
\frac{\dd n^{}_i}{\dd t} + 3 H n^{}_i 
\Eq
- {}^{} \Delta N^{i}_{ijk \to lm} \, g^{}_l {}^{} g^{}_m
\langle ijk \to lm \rangle
\nn &&
\times 
\sx{1.2}{\bigg(}  \hs{-0.05cm}
n^{}_i n^{}_j n^{}_k - n^{}_l n^{}_m
\frac{n^\tf{eq}_i n^\tf{eq}_j n^\tf{eq}_k}{n^\tf{eq}_l n^\tf{eq}_{m\white{l}}} 
\sx{1.2}{\bigg)} 
\nn[0.1cm] &&
- {}^{} \Delta N^{i}_{ij \to lm} \, g^{}_l {}^{} g^{}_m
\langle ij \to lm \rangle
\nn[0.1cm] &&
\times 
\sx{1.2}{\bigg(}  \hs{-0.05cm}
n^{}_i n^{}_j - n^{}_l n^{}_m
\frac{n^\tf{eq}_i n^\tf{eq}_j}{n^\tf{eq}_l n^\tf{eq}_{m\white{l}}} 
\sx{1.2}{\bigg)} 
~,
\end{eqnarray}
where $n^{}_i\,(n^\tf{eq}_i)$ is the (equilibrium) number density of the species $i$, $\Delta N^{i}_{ijk \to lm}\,(\Delta N^{i}_{ij \to lm})$ is the number difference of the species $i$ in the initial and final states of the $3 \to 2\,\,(2 \to 2)$ process, and $\langle ijk \to lm \rangle$ and $\langle ij \to lm \rangle$ are the thermally-averaged annihilation cross sections of the processes, $ijk \to lm$ and $ij \to lm$, respectively.\,\,Note that we adopt the convention in \cite{Kolb:1990vq}, where the squared matrix element in the cross section is averaged over initial and final spins, and includes the  symmetry factors for identical particles in the initial or final states.


Using the standard procedure for \eqref{nBE}, the full Boltzmann equations for Model B in terms of the comoving number yields are written as  
\begin{eqnarray}
&&\frac{\dd Y^{}_X}{\dd x} 
\nn
&&=
-\frac{s^2}{H x}
\nn
&&\hs{0.3cm}
\times \hs{-0.05cm} 
\Bigg\{
12 \langle \XXXNbNa \rangle \hs{-0.03cm}
\sx{1.2}{\bigg[} 
Y^3_X 
- Y^{}_{N_1} Y^{}_{N_2} \frac{(Y^\textsf{eq}_X)^3}{Y^\tf{eq}_{N_1}Y^\tf{eq}_{N_2}} 
\sx{1.2}{\bigg]}
\nn
&&\hs{1cm}
{+}\,2\langle \XXNbXNa \rangle
\sx{1.1}{\bigg(} \hs{-0.03cm} 
Y^2_X Y^{\white{q}}_{N_2} \hs{-0.05cm}
- Y^{\white{q}}_X Y^{\white{q}}_{N_1} \frac{Y^\tf{eq}_X Y^\tf{eq}_{N_2}}{Y^\tf{eq}_{N_1}} 
\sx{1.1}{\bigg)}
\nn 
&&\hs{1cm}
{+}\,2\langle \XXNaXNb \rangle 
\sx{1.1}{\bigg(} \hs{-0.03cm} 
Y^2_X Y^{\white{2}}_{N_1} \hs{-0.05cm}
- Y^{\white{q}}_X Y^{\white{q}}_{N_2} \frac{Y^\tf{eq}_X Y^\tf{eq}_{N_1}}{Y^\tf{eq}_{N_2}}
\sx{1.1}{\bigg)}
\nn
&&\hs{1cm}
{-}\,\langle \XNbNaXX \rangle 
\sx{1.1}{\bigg(} \hs{-0.03cm} 
Y^{}_X Y^{\white{2}}_{N_2} Y^{\white{2}}_{N_1} \hs{-0.05cm}
- Y^2_X \frac{Y^\tf{eq}_{N_2} Y^\tf{eq}_{N_1}}{Y^\tf{eq}_X}
\sx{1.1}{\bigg)}
\hs{-0.03cm}
\Bigg\} 
\nn
&&\hs{0.4cm}
{-}\frac{s}{H x}
\nn
&&\hs{0.3cm}
\times \hs{-0.05cm} 
\Bigg\{
4\langle \XXNcNc \rangle \hs{-0.03cm}
\sx{1.2}{\bigg[} 
Y_X^2 
- Y_{N_{1,2}}^2 \frac{(Y^\tf{eq}_X)^2}{(Y^\tf{eq}_{N_{1,2}})^2} 
\sx{1.2}{\bigg]} 
\nn
&&\hs{1cm}
{-}\,\langle \NcNcXX \rangle \hs{-0.03cm}
\sx{1.2}{\bigg[} 
Y_{N_{1,2}}^2 
- Y_X^2 \frac{(Y^\tf{eq}_{N_{1,2}})^2}{(Y^\tf{eq}_X)^2} 
\sx{1.2}{\bigg]} 
\hs{-0.03cm}
\Bigg\}
~,
\end{eqnarray}
\begin{eqnarray}
&&\frac{\dd Y^{}_{N_1}}{\dd x} 
\nn
&&=
-\frac{s^2}{H x}
\nn
&&\hs{0.3cm}
\times \hs{-0.05cm} 
\Bigg\{
\,2\langle \XXNaXNb \rangle 
\sx{1.1}{\bigg(} \hs{-0.03cm} 
Y^2_X Y^{}_{N_1} \hs{-0.05cm} 
- Y^{}_X Y^{}_{N_2} \frac{Y^\tf{eq}_N Y^\tf{eq}_{N_1}}{Y^\tf{eq}_{N_2}} 
\sx{1.1}{\bigg)}
\nn
&&\hs{1cm}
{+}\,\langle \XNbNaXX \rangle
\sx{1.1}{\bigg(} \hs{-0.03cm} 
Y^{}_X Y^{}_{N_2} Y^{}_{N_1} \hs{-0.05cm} 
- Y_X^2 \frac{Y^\tf{eq}_{N_2} Y^\tf{eq}_{N_1}}{Y^\tf{eq}_X} 
\sx{1.1}{\bigg)}
\nn 
&&\hs{1cm}
{-}\,4\langle \XXXNbNa \rangle \hs{-0.03cm}
\sx{1.2}{\bigg[} 
Y^3_X 
- Y_{N_1} Y_{N_2} \frac{(Y^\tf{eq}_X)^3}{Y^\tf{eq}_{N_1} Y^\tf{eq}_{N_2}} 
\sx{1.2}{\bigg]}
\nn 
&&\hs{1cm}
{-}\,2\langle \XXNbXNa \rangle
\sx{1.1}{\bigg(} \hs{-0.03cm} 
Y^2_X Y^{}_{N_2} \hs{-0.05cm}  
- Y^{}_X Y^{}_{N_1} \frac{Y^\tf{eq}_X Y^\tf{eq}_{N_2}}{Y^\tf{eq}_{N_1}} 
\sx{1.1}{\bigg)}
\hs{-0.03cm}
\Bigg\}
\nn 
&&\hs{0.4cm}
{-}\frac{s}{H x}
\nn
&&\hs{0.3cm}
\times \hs{-0.05cm} 
\Bigg\{
\langle \NaNaXX \rangle \hs{-0.03cm}
\sx{1.2}{\bigg[} 
Y_{N_1}^2 \hs{-0.05cm}
- Y_X^2 \frac{(Y^\tf{eq}_{N_1})^2}{(Y^\tf{eq}_X)^2} 
\sx{1.2}{\bigg]} 
\nn
&&\hs{1cm}
{-}\,4\langle \XXNaNa \rangle \hs{-0.03cm}
\sx{1.2}{\bigg[} 
Y_X^2 
- Y_{N_1}^2 \frac{(Y^\tf{eq}_X)^2}{(Y^\tf{eq}_{N_1})^2} 
\sx{1.2}{\bigg]} 
\nn
&&\hs{1cm}
{+}\,4\langle \NaNaNbNb \rangle \hs{-0.03cm}
\sx{1.2}{\bigg[} 
Y_{N_1}^2 \hs{-0.05cm}
- Y_{N_2}^2 \frac{(Y^\tf{eq}_{N_1})^2}{(Y^\tf{eq}_{N_2})^2} 
\sx{1.2}{\bigg]} 
\nn
&&\hs{1cm}
{-}\,4\langle \NbNbNaNa \rangle \hs{-0.03cm}
\sx{1.2}{\bigg[} 
Y_{N_2}^2 \hs{-0.05cm}
- Y_{N_1}^2 \frac{(Y^\tf{eq}_{N_2})^2}{(Y^\tf{eq}_{N_1})^2} 
\sx{1.2}{\bigg]} 
\hs{-0.03cm} 
\Bigg\}
~,
\label{dYN1} 
\end{eqnarray}
and $\dd Y_{N_2}/\dd x$ can be obtained by exchanging the right-hand side of Eq.\,\eqref{dYN1} as, $N^{}_{1,2} \to \bar{N}^{}_{2,1}, Y^{}_{N_{1,2}} \to Y^{}_{N_{2,1}}$, and $Y^\tf{eq}_{N_{1,2}} \to Y^\tf{eq}_{N_{2,1}}$, where the $3 \to 2$ and $2 \to 2$ cross sections with the degenerate masses are computed as
\begin{eqnarray}
&&\hs{-0.5cm}
\langle \XXXNbNa \rangle
=
\frac{\sqrt{5} {}^{}{}^{} c^2 (m^{}_X/\Lambda)^4}{2304{}^{}\pi{}^{}m_X^5} 
\bigg( \hs{-0.03cm} 5 + \frac{18}{x}+\frac{12}{x^2} \bigg)
~,
\\[0.1cm]
&&\hs{-0.5cm}
\langle \XXNbXNa \rangle
=
\langle \XXNaXNb \rangle 
\nn[0.1cm]
&&\hs{-0.5cm}
\white{\langle \XXNaXNb \rangle}
=
\frac{\sqrt{5} {}^{}{}^{} c^2 (m^{}_X/\Lambda)^4}{768{}^{}\pi{}^{}m_X^5} 
\bigg( \hs{-0.03cm} 5 + \frac{6}{x}+\frac{4}{x^2} \hs{-0.04cm} \bigg)
~,\quad
\\[0.1cm]
&&\hs{-0.5cm}
\langle \XNbNaXX \rangle
=
\frac{\sqrt{5} {}^{}{}^{} c^2 (m^{}_X/\Lambda)^4 }{384{}^{}\pi{}^{}m_X^5} 
\bigg( {}^{} \frac{3}{x}+\frac{2}{x^2} \hs{-0.05cm} \bigg)
~,
\\[0,1cm]
&&\hs{-0.5cm}
\langle X\hs{-0.03cm}\bar{X} \hs{-0.08cm} \to \hs{-0.08cm} N^{}_k\hs{-0.01cm}\bar{N}^{}_k \rangle
=
\frac{c^4 (m^{}_X/\Lambda)^4}{16{}^{}\pi{}^{}(4\pi)^8m_X^2} {}^{}{}^{}
{\cal I}_\Lambda^{{}^{}2} \hs{-0.05cm} 
\sx{0.9}{\bigg(} 
\frac{\Lambda}{m^{}_X} \hs{-0.02cm} 
\sx{0.9}{\bigg)}
\mathord{\raisebox{0.5\depth}{$\sqrt{1-r_k^2}$}}
\nn
&&\hs{2.1cm}
\times \hs{-0.1cm} \bigg[1-r_k^2 + \frac{3}{4 {}^{} x} \big( 5{}^{}r_k^2-2 \big) \bigg]
~,
\\[0.1cm]
&&\hs{-0.5cm}
\langle N^{}_k\hs{-0.02cm}\bar{N}^{}_k \hs{-0.08cm} \to \hs{-0.08cm} X\hs{-0.03cm}\bar{X} \rangle
=
\frac{3{}^{}c^4 (m^{}_X/\Lambda)^4}{32{}^{}\pi{}^{}(4\pi)^8 x{}^{}{}^{}m_X^2} {}^{}{}^{}
{\cal I}_\Lambda^{{}^{}2} \hs{-0.05cm} 
\sx{0.9}{\bigg(}
\frac{\Lambda}{m^{}_X} \hs{-0.02cm} 
\sx{0.9}{\bigg)}
\frac{\sqrt{r^2_k-1}}{r^{}_k} 
\,,
\\[0.1cm]
&&\hs{-0.5cm}
\langle N^{}_i\hs{-0.01cm}\bar{N}^{}_i \hs{-0.08cm} \to \hs{-0.08cm} N^{}_j\hs{-0.01cm}\bar{N}^{}_j \rangle
=
\frac{c^4 (m^{}_X/\Lambda)^4}{192 {}^{} \pi {}^{} (4\pi)^8 x {}^{}{}^{} m_X^2} {}^{}{}^{}
{\cal I}_\Lambda^{{}^{}2} \hs{-0.05cm} 
\sx{0.9}{\bigg(} 
\frac{\Lambda}{m^{}_X} \hs{-0.02cm} 
\sx{0.9}{\bigg)}
\sqrt{1-\frac{r_j^2}{r_i^2}}
\nn
&&\hs{2.2cm}
\times \hs{-0.1cm} \bigg[r_i^2 - r_j^2 + \frac{5}{4 {}^{} x} \big(4{}^{}r_i^2 - r_j^2\big) \bigg]
~,
\end{eqnarray}
with $k = 1,2,\,(i,j) = (1,2),(2,1)$, and $r^{}_k = m^{}_{N_k}/m^{}_X$.




\end{document}